\documentstyle[prl,multicol,aps,psfig]{revtex}

\begin{document}
\title{Relaxation and Zeno effect in qubit measurements}
\author{S. A. Gurvitz$^{1,2}$ {\rm ,} L. Fedichkin$^3$ {\rm ,}
D. Mozyrsky$^2$ {\rm ,} G. P. Berman$^2$\\
$^1$Department of Particle Physics, Weizmann Institute of Science,
Rehovot 76100, Israel\\
$^2$Theoretical Division, Los Alamos National Laboratory, Los Alamos, NM 87545, US\\
$^3$Department of Physics, Clarkson University, Potsdam, New York
13699-5720, US
}
\maketitle
\begin{abstract}
We consider a qubit interacting with its environment and continuously
monitored by a detector represented by a point contact. Bloch-type
equations describing the entire system of the qubit, the environment 
and the detector are derived. Using these equations we evaluate 
the detector current and its noise
spectrum in terms of the decoherence and relaxation rates of the qubit.
Simple expressions are obtained that show how these quantities can be
accurately measured. We demonstrate that due to interaction
with the environment, the measurement can never localize a qubit even
for infinite decoherence rate.
\end{abstract}
\hspace{1.5 cm}
\begin{multicols}{2}
An account of decoherence and relaxation in quantum evolution of a
two-level system (qubit), interacting with an environment and a
measurement device, has become a problem of crucial importance in
quantum computing. Numerous publications have appeared on this subject
dealing with interactions either with a measurement device
(detector)\cite{g1,kor,but} or with the environment 
(a thermal bath)\cite{leg,shnir}. Generally, the simultaneous
influence of an environment and a detector on a qubit 
is very important for understanding qubit measurements  
because the environment and the detector act on the qubit in different 
ways. For instance, the environment at zero temperature 
relaxes the qubit to its ground state. As a result the qubit  
finally appears in a pure state, even though it was initially in 
a statistical mixture. On the other hand, the measurement device 
puts the qubit in a statistical mixture, even if it was initially 
in a pure state.

One of the most striking measurement effects in which the role of
relaxation has not been investigated is the so-called  Zeno
paradox\cite{zeno}. It consists of total freezing of a qubit in
the limit of continuous measurement. Usually, it is associated
with the projection postulate in the theory of quantum
measurements. Indeed, it follows from the Schr\"odinger equation
that the probability of a quantum transition from an initially
occupied state of a qubit is $P(\Delta t)=a(\Delta t)^2$, where
$a$ is a factor which depends on the system\cite{zeno}. If we assume
that $\Delta t$ is the measurement time which determines the
timescale on which the system is projected into the initial state,
then after $N$ successive measurements the probability of finding
the qubit in its initial state, at time $t=N\Delta t$, is
$P(t)=[1-a(\Delta t)^2]^{(t/\Delta t)}$. Thus $P(t)\to 1$ for
$\Delta t\to 0$, $N\to\infty$ and $t$=const. Including the environment 
into the Schr\"odinger equation for the entire system one would expect  
from the above arguments that the relaxation processes
could only affect the coefficient $a$, but cannot destroy 
the qubit localization in the limit of $\Delta t\to 0$.

This conclusion, however, is not correct. We demonstrate 
in this Letter that any weak relaxation
delocalizes the qubit even in the limit of continuous measurement.
It is shown by using new Bloch-type quantum rate equations
for the description of a qubit interacting with a detector and its
environment. These rate equations are derived from the microscopic
Schr\"odinger equation for the entire system. 
Using these equations, we determine the
qubit behavior without any phenomenological parameters. In
addition we are able to relate the qubit behavior to the
detector outcome and therefore establish how decoherence and
relaxation rates of the qubit can be accurately measured.

With respect to the contradiction of our results to the Zeno paradox
arguments, it is possible that the simple derivation described above 
is not valid when the environment is taken into account.  
One reason is that the expansion  of $P(t)=at^2+\cdots$ near 
$t=0$ is not valid because of the
discontinuity of $P'(t)$ at $t=0$. The latter arises
due to the transition to the continuous and sufficiently flat spectrum  
of the environment\cite{gur1,gur2}. We present
our result for an environment at zero temperature, where the
interplay between decoherence and relaxation is most pronounced.
(The detector, however is far from equilibrium.)

Let us consider the generic example of electrostatic qubit
measurements. The qubit is an electron in a double-dot
system, whereas the detector is a point contact
placed near one of the dots, Fig.~1. When the electron occupies
the first well close to the point contact, Fig.~1a,
the current is smaller than in Fig.~1b due to the electron 
electrostatic repulsion. Thus, the
electron is continuously monitored by the tunneling current. The
entire system can be described by the tunneling Hamiltonian
$H=H_0+H_{PC}+H_{int}$, where
\begin{equation}
  H_0=E_1a_1^\dagger a_1+E_2a_2^\dagger a_2-\Omega_0(a_1^\dagger a_2
  +a_2^\dagger a_1)
  \label{a1}
  \end{equation}
is the qubit Hamiltonian
and $H_{PC},\, H_{int}$ describe the point contact detector and
its interaction with the qubit
\begin{eqnarray} \label{a2}
H_{PC}&=&\sum_lE_la_l^\dagger a_l+\sum_rE_ra_r^\dagger a_r
+\sum_{l,r}(\Omega_{lr}a_r^\dagger a_l+H.c.)
\nonumber\\
H_{int}&=&\sum_{l,r}\delta\Omega_{lr}a^\dagger_2a_2
(a_r^\dagger a_l+H.c.).
\end{eqnarray}
Here $a_l^\dagger (a_l)$ and $a_r^\dagger (a_r)$ are the creation 
(annihilation) operators in the
left and the right reservoirs, and $\Omega_{lr}$ is the hopping 
amplitude between the states $l$
and $r$ of the reservoirs. For simplicity we consider electrons 
as spinless fermions and assume each
of the reservoirs is at zero temperature. The interaction term $H_{int}$ 
generates a change in the
hopping amplitude, $\delta\Omega_{lr}={\Omega'}_{lr}-\Omega_{lr}$. 
We assume that the hopping amplitude is weakly dependent on the
states $l,r$, so that it can be replaced by its average value,
$\Omega_{lr}\simeq\bar\Omega$ and $\delta\Omega_{lr}\simeq
\delta\bar\Omega$. Thus the detector current is
$I_1=e2\pi\bar\Omega^2\rho_L\rho_RV$ when the electron occupies
the first dot, and $I_2=e2\pi (\bar\Omega+\delta\bar\Omega
)^2\rho_L\rho_RV$ when the electron occupies the second
dot\cite{g1}. Here $\rho_{L,R}$ are the density of states in the
reservoirs and $V=\mu_L-\mu_R$ is the bias voltage.
%%%%%%%%%%%%%%%%%%%%%%%%%%%%%%%%%%%%%%%%%%%%%%%%%%%
\begin{figure}
{\centering{\psfig{figure=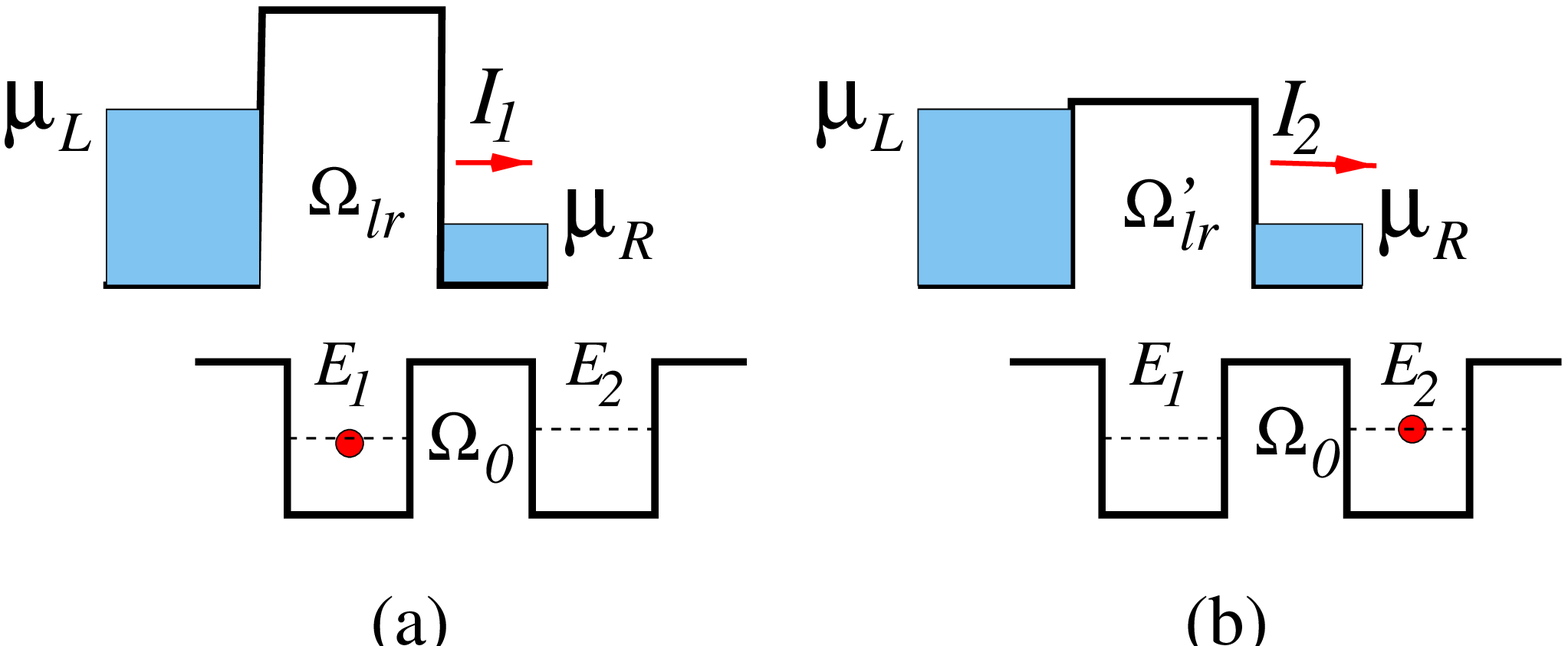,height=4.0cm,width=8.3cm,angle=0}}} {\bf Fig.~1:}
A point-contact detector monitoring the electron position in the double
dot. $\mu_{L,R}$ denote the chemical potentials in the left and right
reservoirs.
\end{figure}
%%%%%%%%%%%%%%%%%%%%%%%%%%%%%%%%%%%%%%%%%%%%%%%%%%%%

It was shown in\cite{g1} that for the case of a large bias voltage
$V$, one can reduce the Schr\"odinger equation for the entire
system to Bloch-type rate equations describing the reduced
density-matrix of the electron, $\sigma_{ij}(t)$. The diagonal
terms of this density matrix, $\sigma_{11}(t)$ and
$\sigma_{22}(t)$ are the probabilities of finding the electron in
the first dot or in the second dot, respectively.
The off-diagonal matrix elements (``coherences''),
$\sigma_{12}(t),\,\sigma_{21}(t)$ describe linear  
superpositions of these states. One finds\cite{g1}
\begin{mathletters}
  \label{a3}
\begin{eqnarray}\label{a3a}
&&\dot\sigma_{11}=-i\Omega_0(\sigma_{12}-\sigma_{21})\\
&&\dot\sigma_{12}=i\epsilon\sigma_{12}-i\Omega_0(2\sigma_{11}-1)
-(\Gamma_d/2)\sigma_{12}\, , \label{a3b}
\end{eqnarray}
\end{mathletters}
where $\sigma_{22}(t)=1-\sigma_{11}(t)$ and
$\sigma_{21}=\sigma^*_{12}(t)$. Here $E_{1,2} =\mp\epsilon/2$ and
$\Gamma_d =\left(\!\sqrt{I_1/e}-\sqrt{I_2/e}\,\right)^2\!V/2\pi$
is the decoherence rate due to interaction with the detector.
Solving Eqs.~(\ref{a3a},\ref{a3b}) one obtains that the
decoherence term in Eq.~(\ref{a3b}) leads to a vanishing of
coherences in the limit $t\to\infty$. Finally the electron
density matrix becomes a completely random mixture for {\em any}
initial condition: $\sigma_{ij}(t\to\infty )=(1/2)\delta_{ij}$.

Now we introduce the environment, represented by a boson bath at
zero temperature and interacting with the electron. First,
consider the case in which the electron is not coupled to the
detector, $H_{int}=0$. Then, for any initial conditions, the
electron relaxes to its lowest energy state ($E_+$) through 
boson emission. This final state can be found by disregarding any 
possible renormalization of the electron states due to interaction
with the boson field. We thus diagonalize the Hamiltonian $H_0$,
Eq.~(\ref{a1}) by using the following ``rotation'':
$a_{+,-}=\pm \cos (\theta /2)a_{1,2}+\sin (\theta /2 )a_{2,1}$,
with $\tan\theta =2\Omega_0/\epsilon$.
Then $H_0=(\tilde\epsilon /2) (a_-^\dagger a_--a_+^\dagger a_+)$, where
$\tilde\epsilon =(\epsilon^2+4\Omega_0^2)^{1/2}$.

In order to account for the above relaxation effects in the qubit
evolution, we replace the Hamiltonian $H_0$ by the following
Hamiltonian:
\begin{eqnarray}
H'_0=H_0 +\sum_\alpha \big [E_\alpha b^\dagger_\alpha b_\alpha
+c_\alpha
  (a^\dagger_+a_-b_\alpha^\dagger +a^\dagger_-a_+b_\alpha )\big ] ,
  \label{a5}
\end{eqnarray}
where  $b^\dagger_\alpha (b_\alpha)$ are creation (annihilation)
operators of a boson with the energy $E_\alpha$. This Hamiltonian
is essentially equivalent to the Lee model and has been
investigated in numerous works\cite{pfeifer}. In the weak coupling
limit this model leads to the same results as the usual spin-boson
model, although it includes an additional, direct coupling between 
the qubit states $a^\dagger_{1,2}|0\rangle$.

If there is no interaction with the detector, $H_{int}=0$, and
$c_\alpha$ is weakly dependent on $E_\alpha$, one can trace
the boson variables in the Schr\"odinger equation, by reducing it
to the rate equations for the electron density matrix in the basis
states, $a^\dagger_{\pm}|0\rangle$\cite{pfeifer}
\begin{mathletters}
\label{a6}
\begin{eqnarray}
  && \dot\sigma_{--}(t)=-\Gamma_r\sigma_{--}(t)
  \label{a6a}\\
  && \dot\sigma_{+-}(t)=i\tilde\epsilon\sigma_{+-}(t)
  -(\Gamma_r/2)\sigma_{+-}(t)\, ,\label{a6b}
 \end{eqnarray}
\end{mathletters}
\noindent with $\sigma_{++}(t)=1-\sigma_{--}(t)$ and
$\sigma_{-+}(t)=\sigma^*_{+-}(t)$. Here
$\Gamma_r=2\pi c_\alpha^2\rho_\alpha$ is the relaxation rate
and $\rho_\alpha$ is the density of boson states.
Eqs.~(\ref{a6a},\ref{a6b}) obviously reproduce an exponential
decay (relaxation) of the electron from the upper level $E_-$ to
the ground level $E_+$: $\sigma_{--}(t)=\exp (-\Gamma_r t)$. Note
that the off-diagonal density-matrix element $\sigma_{+-}(t)$
vanishes in the limit $t\to\infty$, similar to $\sigma_{12}(t)$ in
Eq.~(\ref{a3b}). Yet, the disappearance of off-diagonal density
matrix elements does not necessarily imply dephasing. Indeed, in
the case of relaxation, Eqs.~(\ref{a6a},\ref{a6b}), the diagonal
term $\sigma_{--}(t)$ vanishes as well for $t\to\infty$. As a
result, the qubit finally appears in a pure state (the ground
state), in contrast to Eqs.~(\ref{a3a},\ref{a3b}) leading to the
statistical mixture. Now let us include the interaction with the
detector. Then the Hamiltonian of the entire system becomes
$H'=H'_0+H_{PC}+H_{int}$. It is useful to return to the original
qubit basis $a^\dagger_{1,2}|0\rangle$,
in which $H_{int}$, Eq.~(\ref{a2}),
has a simple form. The corresponding rate equations for the qubit
density matrix $\sigma_{ij}(t)$ are obtained by tracing the
detector and boson degrees of freedom. These equations can be
written directly by using the method of Refs.\cite{g1,gp}. We
obtain
\begin{mathletters}
\label{a9}
\begin{eqnarray}\label{a9a}
&&\dot\sigma_{11}=-i\Omega_0(\sigma_{12}-\sigma_{21})
-\Gamma_r(\kappa\epsilon /2 \tilde\epsilon)
(\sigma_{12}+\sigma_{21})
\nonumber\\
&&~~~~~~~~~~~~-(\Gamma_r/4)[1+(\epsilon/\tilde\epsilon )^2](2\sigma_{11}-1)
+\Gamma_r(\epsilon /2\tilde\epsilon)
 \\[5pt]
  &&\dot\sigma_{12}=i\epsilon\sigma_{12}
  -[i\Omega_0+\Gamma_r(\kappa\epsilon /2\tilde\epsilon)]
  (2\sigma_{11}-1)
\nonumber\\
&&~~+\Gamma_r[\kappa -(1/2)\sigma_{12}
  -\kappa^2(\sigma_{12}+\sigma_{21})]
-(\Gamma_d/2)\sigma_{12}\, ,
\label{a9b}
\end{eqnarray}
\end{mathletters}
where $\kappa =\Omega_0/\tilde\epsilon$. (Similar equations 
were obtained by Korotkov~\cite{kor1} in the weak coupling limit 
by using phenomenological arguments). 

Solving Eqs.~(\ref{a9a},\ref{a9b}) for $\epsilon =0$, one obtains
the qubit density matrix for the stationary state ($\dot\sigma =0$)
\begin{equation}
\bar\sigma =\sigma(t\to\infty ) = \pmatrix{1/2&y/(1+2y)\cr
y/(1+2y)& 1/2} \label{a10}
\end{equation}
where $y=\Gamma_r/\Gamma_d$. This describes a heated qubit with an
effective temperature $T_{eff}=2\Omega_0/\ln (1+4y)$. This heating
is caused by the measurement process\cite{moz}.

Now, we investigate how the relaxation affects the time-dependence
of the qubit density matrix. Consider again the symmetric case,
$\epsilon =0$. Let us evaluate the probability of finding the
electron in the first dot, $\sigma_{11}(t)$. By solving
Eqs.~(\ref{a9}) for the initial conditions $\sigma_{11}(0)=1$,
$\sigma_{12}(0)=0$. We find
\begin{equation}
\sigma_{11}(t)=\frac{1}{2}
+{e^{-\Gamma_rt/2}\over 4}\left (C_1
e^{-e_-t}
+C_2e^{-e_+t}\right )
\label{a15}
\end{equation}
where $e_{\pm}={1\over4}(\Gamma_d\pm\Omega)$, $\Omega
=\sqrt{\Gamma_d^2-64\Omega_0^2}$ and
$C_{1,2}=1\pm(\Gamma_d/\Omega)$. Thus, for the case of weak
decoherence, $\Gamma_d\ll 8\Omega_0$, the electron displays damped
oscillations between the dots with the Rabi frequency, 
$\sqrt{(2\Omega_0)^2-(\Gamma_d/4)^2}$. 

For strong coupling to the detector, $\Gamma_d\gg
8\Omega_0$, the situation is different. If $\Gamma_r=0$, the
electron would stay in the same dot for a long time (``quantum
Zeno'' effect). The  dwell time $\tau_Z$ obtained from
Eq.~(\ref{a15}) is $\Gamma_d/8\Omega_0^2$. Thus the increase of
$\Gamma_d$ leads to a freezing of the electron, which is totally
localized in the limit of $\Gamma_d\to\infty$. This result
is consistent with the Zeno paradox, based on the projection postulate.
Indeed, the ``measurement time'' $\Delta t$ is inversely
proportional to $\Gamma_d$. (See for instance\cite{but,sh}). 
Also one can observe from Eq.~(\ref{a15}) that for small $\Delta t$,  
$P(\Delta t)=[1-\sigma_{11}(t)]\propto (\Delta t)^2$. 
However, as follows from Eq.~(\ref{a15}), the interaction with the
environment essentially destroys the Zeno effect. We find that the
Zeno time in this case is
$\tau_Z^{-1}=(\Gamma_r/2)+(8\Omega_0^2/\Gamma_d)$. Therefore the
continuous measurement cannot localize the electron for a long
time, even if $\Gamma_d\to\infty$. The corresponding dwell time is
restricted by the relaxation rate $\Gamma_r^{-1}$. The disappearance 
of Zeno paradox can be understood as follows. It is crucial to observe 
that the relaxations due to the environment and the 
detector are factorizable, as seen in 
Eq.~(\ref{a15}). As a result of this factorizability, an effective 
linear in $\Delta t$ term, generated by purely exponential decay for small 
$\Delta t$, appears in the expansion of $P(\Delta t)$. This leads to 
the elimination of the Zeno effect\cite{gur1,gur2}. 

Now, we will relate the qubit behavior to the
corresponding observable quantities. For this, we have to
include the detector states in the rate equations $(\ref{a9})$. We
thus introduce the (reduced) density matrix $\sigma_{ij}^{(nn')}(t)$, where
index $n$ denotes the number of electrons that have arrived in the
right reservoir by the time, $t$\cite{gur2}. This density matrix is related to
the previous one by $\sigma_{ij}(t)=\sum_n\sigma_{ij}^{(n)}(t)$,
where $\sigma_{ij}^{(n)}\equiv \sigma_{ij}^{(nn)}$. Starting from
the microscopic Schr\"odinger equation for the entire system and
using the same method as in Refs.\cite{g1,gur2,gp} we can demonstrate
that in the limit of high bias-voltage $V$ of the detector,
Fig.~1, the off-diagonal density-matrix elements
$\sigma_{ij}^{(nn')}$ are decoupled from the diagonal elements
$\sigma_{ij}^{(n)}$ in the equation of motion\cite{gur2}. As a result we
arrive to the following Bloch-type rate equations:
\begin{mathletters}
\label{a11}
\begin{eqnarray}
&&\dot\sigma_{11}^{(n)}=
-i\Omega_0(\sigma_{12}^{(n)}-\sigma_{21}^{(n)})
-\Gamma_r(\kappa\epsilon /2\tilde\epsilon)(\sigma_{12}^{(n)}+\sigma_{21}^{(n)})
\nonumber\\
&&-{\Gamma_r\over4}
(\beta_-^2\sigma_{11}^{(n)}
    -\beta_+^2\sigma_{22}^{(n)})
-{I_1\over e}(\sigma_{11}^{(n)}-\sigma_{11}^{(n-1)})
\label{a11a}\\[5pt]
&&\dot\sigma_{22}^{(n)}=
i\Omega_0(\sigma_{12}^{(n)}-\sigma_{21}^{(n)})
+\Gamma_r(\kappa\epsilon /2\tilde\epsilon)(\sigma_{12}^{(n)}+\sigma_{21}^{(n)})
\nonumber\\
&&+{\Gamma_r\over4}
(\beta_-^2\sigma_{11}^{(n)}
    -\beta_+^2\sigma_{22}^{(n)})
-{I_2\over e}(\sigma_{22}^{(n)}-\sigma_{22}^{(n-1)})
\label{a11b}\\[5pt]
&&\dot\sigma_{12}^{(n)}=i\epsilon\sigma_{12}^{(n)}
-i\Omega_0(\sigma_{11}^{(n)}-\sigma_{22}^{(n)})
\nonumber\\
&&+\Gamma_r[\kappa (\beta'_-\sigma_{11}^{(n)}+\beta'_+\sigma_{22}^{(n)})
-(1/2)\sigma_{12}^{(n)}-\kappa^2(\sigma_{12}^{(n)}+\sigma_{21}^{(n)})]
\nonumber\\
&&~~~~~~~~~~~~~~~~~~~~~~~~-{I_1+I_2\over 2e}\sigma_{12}^{(n)}+
 {\sqrt{I_1I_2}\over e}\sigma_{12}^{(n-1)}\, ,
  \label{a11c}
\end{eqnarray}
\end{mathletters}
where $\beta_{\pm}=1\pm (\epsilon/\tilde\epsilon )$ and
$\beta'_\pm=1\pm(\epsilon/2\tilde\epsilon )$. Tracing Eqs.~(\ref{a11})
over $n$ and using $\sigma_{11}(t)+\sigma_{2}(t)=1$ we
obtain Eqs.~(\ref{a9}).

Eqs.~(\ref{a11}) allow us to evaluate the average detector current
and its shot-noise power spectrum. The (ensemble) average current is
given by
\begin{equation}
  I(t) =e\sum n\dot P_n(t) =(I_1-I_2)\sigma_{11}(t)+I_2
  \label{a12}
\end{equation}
where $P_n(t)=\sigma_{11}^{(n)}(t)+\sigma_{22}^{(n)}(t)$ is the probability
of finding $n$ electrons in the collector by time $t$.
As expected, the average current is directly related to the occupation
of the first dot.  The shot-noise power spectrum can be calculated via the
McDonald formula\cite{mac,moz1}
\begin{equation}
S(\omega) = {e^2\omega\over \pi} \int_0^\infty dt
\sin (\omega t) {d\over dt}N_R^2 (t)\, ,
\label{a13}
\end{equation}
where $N_R^2 (t) =\sum_n n^2P_n(t)$.

Now, we investigate how the relaxation and decoherence rates can
be extracted from $I(t)$ and $S(\omega)$. Consider first the
stationary detector current $\bar I = I(t\to\infty )$ for the
symmetric case ($\epsilon =0$). It follows from Eqs.~(\ref{a10})
and (\ref{a12}) that $\bar I=(I_1+I_2)/2$, so that it is
independent of $y=\Gamma_r/\Gamma_d$. Therefore this ratio cannot
be extracted from $\bar I$ for $\epsilon =0$. However, for a non
symmetric qubit, $\epsilon\gg\Omega_0$, the detector current
becomes sensitive to $y$. Indeed, $\bar I =(I_1+I_2)/2$ for
$\Gamma_r=0$ and $\Gamma_d\not =0$ , but $\bar I \simeq I_1$ for
$\Gamma_d=0$ and $\Gamma_r\not =0$ (since the relaxation puts the
system into the lowest energy state, $E_+\sim E_1$ for
$\epsilon\gg\Omega_0$). Using Eqs.~(\ref{a9}) and (\ref{a12}) we
obtain for $\epsilon\gg\Omega_0$
\begin{equation}
\bar I =\Delta I  {\displaystyle y+(\Omega_0/\epsilon )^2\over
    \displaystyle y+2(\Omega_0/\epsilon )^2}+I_2\, ,
  \label{a14}
  \end{equation}
where $\Delta I =I_1-I_2$.

Although in the symmetric case $\epsilon =0$, the relaxation does
not affect the stationary current $\bar I$. It instead affects its
transient properties, which are reflected in the shot-noise
spectrum of the detector current, $S(\omega )$, given by
Eq.~(\ref{a13}). One can write $S(\omega )=S_0+\Delta S(\omega )$,
where the first term $S_0=e(I_1+I_2)$ is the Schottky noise and
the second term is the excess noise generated by the qubit
dynamics. Generally the analytic expression for $\Delta S(\omega
)$ is rather lengthy. We therefore present it only for $\epsilon
=0$ and in two limits: $\Gamma_d ,\Gamma_r\ll\Omega_0$ and
$\Gamma_d \gg\Omega_0$. In the first case the excess noise can be
very well approximated by a Lorentzian
\begin{equation}
  \Delta S(\omega )={(\Delta I)^2(\Gamma_d+2\Gamma_r)\over
    (\Gamma_d+2\Gamma_r)^2+16(\omega -2\Omega_0)^2}
\label{a16}
\end{equation}
This corresponds to the result of Korotkov, obtained by 
a Bayesian approach (``continuous'' wave function collapse) in the weak 
coupling limit\cite{kor1}. 

The second case, $\Gamma_d \gg\Omega_0$, corresponds to the Zeno
effect regime. We find
\begin{equation}
  \Delta S(\omega )={(\Delta I)^2(16\Gamma_d\Omega_0^2+\Gamma_r\Gamma_T^2)
    \over
    4[\Gamma_d^2\omega^2+4(\omega^2-4\Omega_0^2)^2]
    +\Gamma_r\Gamma_d\Gamma_T^2}\, ,
\label{a17}
\end{equation}
where $\Gamma_T=\Gamma_d+\Gamma_r$.
%%%%%%%%%%%%%%%%%%%%%%%%%%%%%%%%%%%%%%%%%%%%%%%%%%%
\begin{figure}
{\centering{\psfig{figure=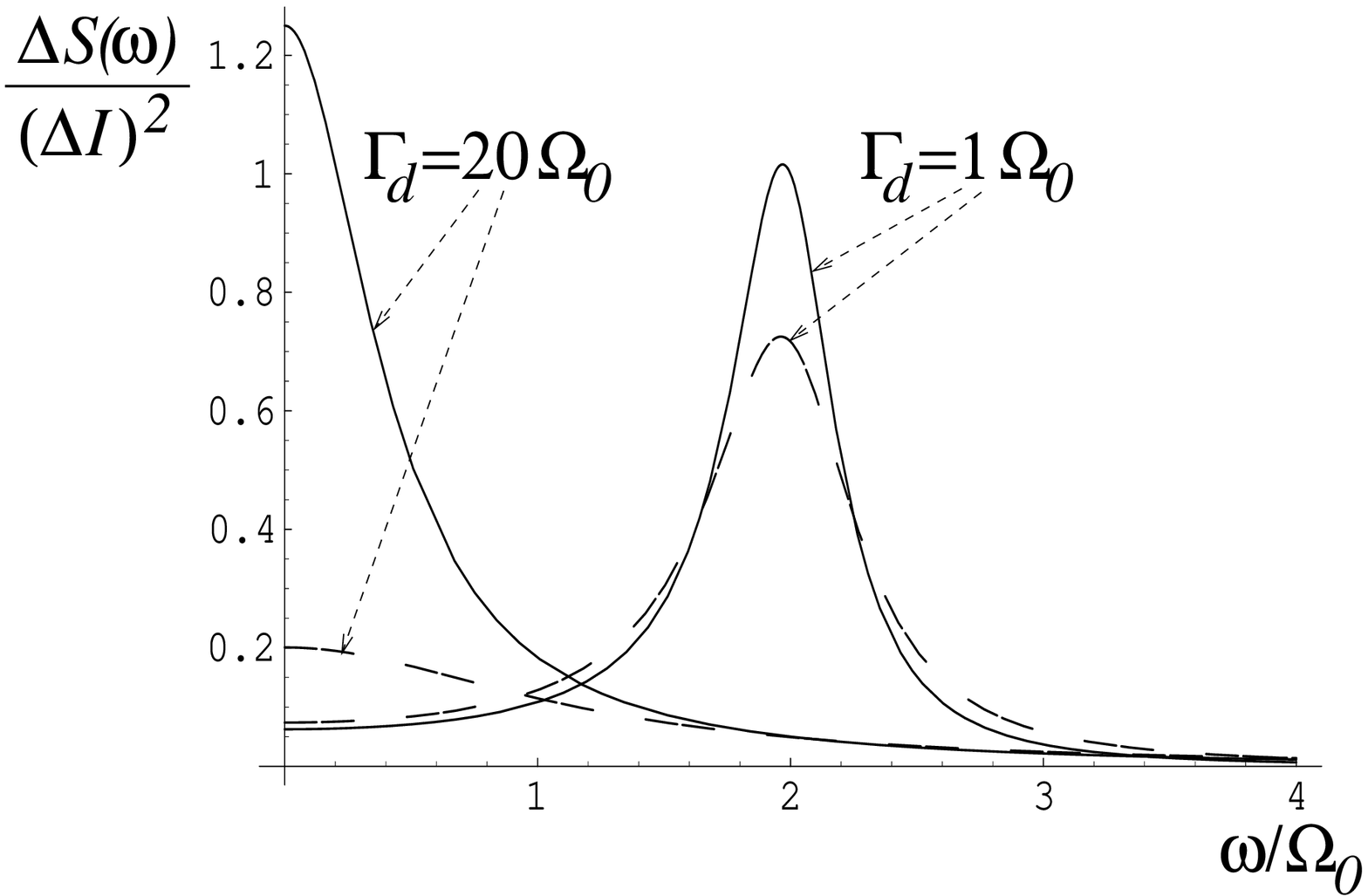,height=4.4cm,width=8.3cm,angle=0}}} {\bf Fig.~2:}
The excess noise power spectrum of the detector current for different
decoherence and relaxation rates. The solid curves correspond to
$\Gamma_r=0$ and the dashed curves to $\Gamma_r=0.1\Omega_0$.
\end{figure}
%%%%%%%%%%%%%%%%%%%%%%%%%%%%%%%%%%%%%%%%%%%%%%%%%%%%

The excess noise, $\Delta S(\omega )$, for different
values of $\Gamma_d ,\Gamma_r$ and $\epsilon =0$ are shown in
Fig.~2. As expected, for $\Gamma_d\lesssim\Omega_0$, the
Rabi oscillations generate a peak in the noise spectrum at $\omega
=2\Omega_0$. The relaxation modifies this peak according to
Eq.~(\ref{a16}). In the case of large decoherence rate
$\Gamma_d\gg\Omega_0$ and $\Gamma_r=0$, the qubit is in the regime
of the Zeno effect. This leads to a telegraph noise\cite{kor1},
resulting in a peak at $\omega =0$. This peak, however, is
strongly diminished in the presence of relaxation, even for a
small $\Gamma_r$, as given by Eq.(\ref{a17}).

This strong dependence provides a new way to measure relaxation
rate of a quantum system in experiments. As is seen in Fig.~2, one
can measure the relaxation rate of a qubit via the noise spectrum of
the detector at zero frequency. Specifically, the relaxation rate can
be lower by two or more orders of magnitude compared with the
dephasing rate and still change the noise spectrum
significantly. With such sensitivity, one may more
accurately measure the relaxation rate because this allows one to increase
the output signal by increasing the coupling between the qubit and
the detector.

In summary, we found that a qubit interacting with its
environment and with a detector can be described by a set of modified
Bloch-type equations in which the decoherence and relaxation
process are clearly distinguished. The most interesting result of 
our analysis is that there is no Zeno paradox when the relaxation 
due to the environment is taken into account. 
In addition, we obtained simple analytical expressions for
the detector current and its noise spectrum. Using these findings, 
we proposed a new and possibly more accurate way to measure 
the qubit decoherence and relaxation rates. 

This work was supported by the US DOE under
contract W-7405-ENG-36, by NSA and ARDA. L.F. and D.M. were
supported, in part, by the US NSF grants ECS-0102500 and
DMR-0121146.

\end{multicols}
\end{document}